# Network planning tool based on network classification and load prediction

Seif Eddine Hammami, Hossam Afifi, Michel Marot, Vincent Gauthier
RST Department
Institute Mines-Telecom, Télécom SudParis
Saclay, France
{seif_eddine.hammami, hossam.afifi, michel.marot, vincent.gauthier }@telecom-sudparis.eu

*Abstract*— Real Call Detail Records (CDR) are analyzed and classified based on Support Vector Machine (SVM) algorithm. The daily classification results in three traffic classes. We use two different algorithms, K-means and SVM to check the classification efficiency. A second support vector regression (SVR) based algorithm is built to make an online prediction of traffic load using the history of CDRs. Then, these algorithms will be integrated to a network planning tool which will help cellular operators on planning optimally their access network.

## I. INTRODUCTION

The wide development of cellular networks and the impressive expansion of mobiles and wireless equipment usage may oblige telecommunication operators to innovate in strategies for resource allocation. Mainly high user mobility and load variation in the day and the week show that new allocation techniques need to be designed, to provide better QoS and face the continuous increase of users demand. The deployment of femtocells [1] or WiFi offloading approaches are potential solutions.

On the other side, mobile phone traces (including CDRs) constitute a source of unexploited information that can be used for many purposes. These CDRs, collected from cellular networks operators, may contain the ID of users, the cell ID, the geographical position of cell towers the users was attached to when an ingoing/outgoing call or message occur and its timestamps. The huge amounts of information which can be extracted from these datasets provide an important contribution to related domains and help in studying human mobility [2] or urban traffic [3]. These traces are also useful for telecommunication operators to analyze their network data traffic [4] and their global efficiency, they can detect potential areas where users consume more traffic and to detect when and where network bottlenecks occur. So, they can find solutions to prevent congestion and optimize their cellular access network to guarantee a better quality of service.

Before rethinking cellular network deployment, operators must begin by analyzing the existing architecture. CDRs can be very useful for this step due to the rich quantity of information that they offer. Operators can then propose new algorithms for the optimization of the actual resource allocation. In addition, an online prediction algorithm made to forecast the load or the number of users attached to each antenna can be very useful as well to build a dynamic set of resource distribution techniques.

In this paper, we will tackle the base station profiles classification and the load prediction of base stations issues. Classification can provide a clear overview of the cellular network and detect all the different BS profiles in the network, and can help us along with the load prediction to enhance radio resources allocation and energy saving. For the classification problem we use two different types of classifiers, a modified unsupervised classifier based on k-means and support vector machine (SVM) which is a supervised algorithm. The 'old' classification tools, like hidden Markov chains, neural networks, etc, consume more processing time, computational resources and need a complex configuration to be implemented. That's why we choose and focus on support vector machine algorithms. Unlike the other machine learning algorithm; after the training step, SVM didn't need to use all the training dataset for classification but it works only with support vectors, which reduces notably the processing time, in addition to its minimal computational requirements. On the other hand, the principles of SVM can be used for other tasks like prediction. So, we decide to continue with Support Vector Regression to predict the traffic load for each base station. All the algorithms presented in this paper use a real set of traces: the CDRs data set provided by the orange D4D challenge [5]. Then, we propose a planning tool based on these algorithms.

The paper is organized as follow, section II present some related works. Section III presents a brief analysis of the traces dataset and the access network of Senegal. In, Section IV we present our classification model and show their results. Section V depicts the prediction model and the obtained results. Then, section VI, will present our network planning tool. Finally, in section VI we conclude the paper and present some future works.

## II. RELATED WORKS

Mobile device activity [4] has become the focus of studies on user's call and/or data traffic information [6, 7], user's mobility [8] besides information about core network traffic. So, this important quantity of information can be exploited in several fields. Recently, some works focus on cellular networks traces and try to exploit them to study human mobility behavior [2], other use it to characterize individual movements [9] and to model them [10]. But basically all these contributions uniquely operate on geographical information provided by the traces and omit temporal data that can be important to study the dynamicity of some phenomena. Some scientists interested in this temporal behavior try to exploit it to find out commuted spatio-temporal patterns [11] and use it to characterize urban traffic systems [3]. Mobile traces analysis can also be helpful for city planning researchers. These datasets can be useful to categorize the cities and understand their structure [12]. Mobile traces are also exploited to infer the most visited sites by users [13] in order to develop tourism applications [14] or to semantically annotate those places [15].

Other studies focus on the cellular network use cases, exploiting traces for networking applications as we also do in this paper. M. Fiore *et al.* [16] have analyzed similar realistic CDRs datasets in order to classify call profiles. For the classification, they proposed a snapshot based framework extracted from the dataset. Authors in [15] classify and annotate the important base stations with a semantic label such as home and work using an algorithm based on machine learning.

In our contribution, we used two types of algorithms to classify the base stations, an unsupervised algorithm: k-means, and the supervised algorithm using Support Vector Machine (SVM). Mainly, these algorithms are used for data mining in signal processing applications such as speech separation [17] or signal modulation classification [18]. They can be used also for image processing [19], moreover in bio-informatics fields to classify for example a set of genes or bacteria [20]. Recently, with the explosion of the internet content, machine learning algorithms like SVM are also used to classify web texts [21], IP traffic [22] or even to detect network intrusion [23].

Support vector machine was initially developed for classification purposes, but it has been extended to deal with prediction tasks. Usually, the support vector regression (SVR) is employed to predict time series [24] like the fluctuation of stock market prices [25] or to predict the maximum atmospheric temperature [26] and to forecast the rainfalls [27]. In [28], the author uses SVR to predict home electricity load peak.

In [29] authors developed a SON (Self Organized Network) algorithm based on SVR to predict the load of eNodeB for the purpose of enhancing the functionality of the scheduler and avoid the degradation of the downlink. In [30], a non-statistical algorithm based on chaos theory and multivariate traffic time series, is proposed to predict the traffic load for a heterogeneous wireless network.

These contributions used a synthetic data, which may not describe properly the real behavior of networks, to validate their prediction algorithm. Since everything is dependent on training, we think that using simulated data drastically deteriorates the benefits of such contributions. In fact, using synthetic data instead of real traces does not allow to work with complex and long term correlations which can affect load predictions. That is why it is crucial, when dealing with load prediction, to use real traces. In our work, the D4D real dataset is used to test our classification and prediction algorithms. Moreover, it is double checked with different datasets from another country. We hence explain hereafter how these algorithms will perform with real traces.

III. ANALYSING D4D TRACES

For our study, we used D4D-Senegal challenge traces [5] to analyze mobile phone traffic and to model our framework of profile classification and load prediction. The D4D traces are based on call detail records of phone calls and short messages exchanges of about 9 million users for the year 2013. The datasets are divided into 3 sets: one set contains the antenna-to-antenna traffic for 1666 antennas on an hourly basis, another contains one year of coarse-grained mobility data at district level and a last one contains fine-grained mobility data on a rolling 2-week basis for a year with bandicoot behavioral indicators at individual level for about 300,000 randomly sampled users. We exploited the last set to analyze the traffic load and the repartition of important base stations (heavy duty) over the country. From these traces we extract important information such as user location according to BS attachment on a specific time. These information allow us to track users, define their most visited places and have an idea about instant traffic load and instant BS capacity for the whole day. That helps us to identify important base stations and infer the profile of each BS.

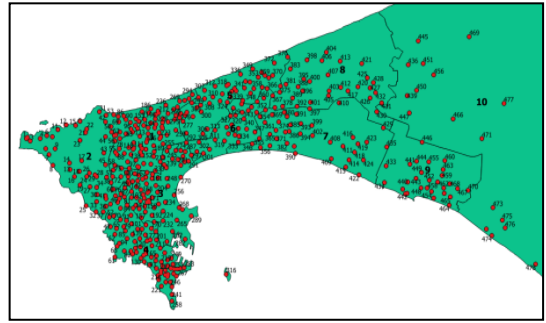

Fig. 1. Dakar Access Network map

This first set of information is extracted by implementing a python framework, dealing with big data analysis: to count the number of users attached to each base station, in each instant and to visualize the repartition of network load over the radio access network implanted on Senegal (Fig 1 shows an example of the access network in Dakar).

Results represented in Fig 2 show the evolution of number of users per BS on three different districts of Dakar during a day (Each line of each plot depends on one BS and each plot represents all the BS of a district).

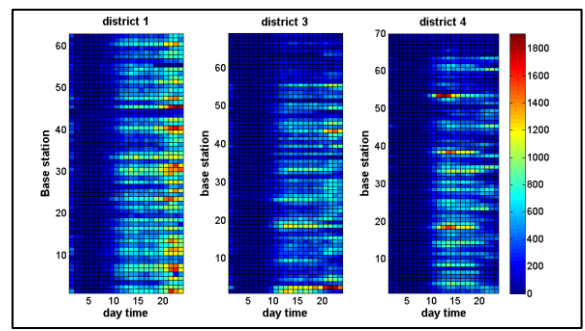

Fig. 2. Load repartition in three districts in Dakar

From these results, we notice that we have almost three dominant BS profiles: Some have a constant activity where the fluctuation of capacity and load is not very important, we call this profile the "always loaded class" (or class1). For some BSs, we noticed that they may have a peak load between 8 am and 2pm and then the load decreases clearly for the rest of the day and there is no more important capacity during the night. We called this profile the "morning peak load class" (or class2). Finally, for the rest of BS, we notice that the load is more important during the night and even late night and we called that profile the "evening peak load class" (or class3).

Based on this analysis, we built our classification model using our three profiles to train the support vector machine classification model applied later to the whole traces dataset.

## IV. CLASSIFICATION OF BASE STATION

### A. SVM Based Classification:

1. Support vector machine classifier:

The Support vector machine (SVM) was proposed by Vapnik in [31] and it was used for computational and statistical learning. Recently, the use of SVM is expanded to several fields like image processing, medical analysis, text classification, etc. In our study, we used SVM to classify the BSs profiles.

The main idea behind SVM is to maximize the margin between classes and minimize a risk error function. Given a data set $\{x_i; y_i\}$; $i = 1,...,m$ where $y_i \in \mathbb{R}$ represents the label of an arbitrary training example $x_i \in \mathbb{R}^d$; $d$ is the dimension of the input space. The SVM defines a linear decision hyperplane, represented by the equation $f(x) = w.x+b$, that separates the training data by maximizing the margin between different classes and the distance to this hyperplane. So we can formulate the primary optimization problem as following:

$$\begin{cases} \text{Min } \frac{1}{2}\|w^2\| + C \sum_1^m \xi_i \\ \text{Sb. to } y_i(w. x_i + b) > 1 - \xi_i \\ \xi_i > 0 \end{cases}$$

where C is a real positive constant and determines the margin between the margin and training error, and $\xi_i$ is a slack parameter.

The solution of this problem is $f(x) = \sum_1^m \alpha_i^* y_i (x.x_i) + w_0^*$, and the training instances that lie closest to the hyperplane are called support vectors.

2. Classification model:

The SVM classification procedure considered in our research is based on the Multi-class Support Vector Machine algorithm proposed by Weston and Watkins in [32].

- Training step:

As known, the first step before building the classification model is the training step. For that purpose, we used a beforehand classified set of some base stations from traces that we deduced their classes during initial analysis. Thus, our training vectors, $x_i$, represent the number of users (load) attached to $BS_i$ over each interval of 10 minutes during the day. Fig 3 shows the training profiles for BSs corresponding to class1, class2 and class3 respectively. Each vector must be trained with the label, $y_i$, of the corresponding class. We choose the RBF kernel function because our training data are non linear and it matches very well with our issue. So, we must normalize our training data before starting the training step. Globally, the training is represented as following:

$$\{(x_1,y_1), (x_2,y_2), .....(x_n,y_n)\}$$

Where $x_i \in \mathbb{R}^n$, (n= 144), represents the training vectors and $y_i \in \{1, 2, 3\}$ represents the class labels.

- Testing step:

After building the classification model, we must ensure that it works well and has good performance to classify all the rest of the base stations. For that, we used also another beforehand known set of BSs to compare its classification results with the reality and we used for this step the cross validation technique to adjust the model with the best parameters.

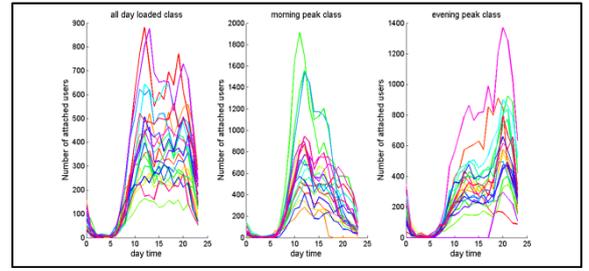

Fig. 3. Training data divided into the appropriate three classes: always loaded class (left), morning peak load class (middle) and evening peak load class (right) (each line belongs to one BS)

### B. K-means Based classification:

To validate the SVM classification model we should compare its performance to another classifier. We choose for that an algorithm based on K-means. K-means is an unsupervised data mining algorithm, proposed by J.B MacQueen [33], which achieves clustering or classifying of data into K different sets.

The k-mean algorithm is able to classify data into a predefined number of clusters so that the distance between the elements of the clusters is minimized, but it is not possible to train it on a predefined dataset, that is why we decide to use it in a two step manner. First, we run the k-means algorithm on the training data under the constraint to obtain three clusters. The trained data are then classified into three reference clusters (fig 4). We observed that we obtain more or less the same classification as the one used to train SVM. After this training phase, the second step consists in assigning a BS to a cluster by calculating its (Euclidean) distance to the nearest average of the three training clusters and then the classification is made.

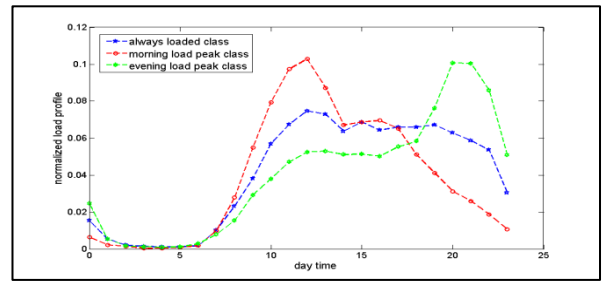

Fig. 4. Normalized average load of three training clusters used for k-means classification.

### C. Results and Disscussion:

The main objective of these models is to have an automatic and daily classification of all the base stations, for all the D4D dataset, according to the number of users attached to during all the day.

We run our classification models on traces to classify base stations installed in one district (Dakar plateau) of Dakar. Each base station is represented by a normalized vector of load at each 10 minutes interval of the day.

Figures 5 depict an overview of the comparison between the results obtained from the classification by k-means (left curves), SVM (middle curves) and real classification (right curves) by plotting the daily load variation of each BS in the appropriate class i.e. the left plot of the fig 5.a aggregates all BSs corresponding to the first class of K-means and each colored line represents one BS. Table II resumes the numerical results of the comparison between different classifications. Results prove that the SVM is lightly better than k-means with an 87.14% correctly classified BSs in total against 85% for K-means and 9 wrongly classified BSs against 10 for K-means.

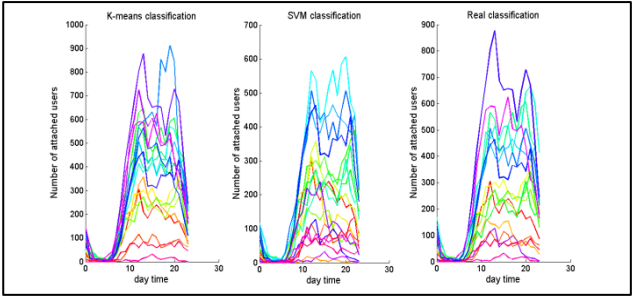

Fig. 5. a. Classification result of the always loaded class, by k-means (left), by SVM (middle) and real classification (right)

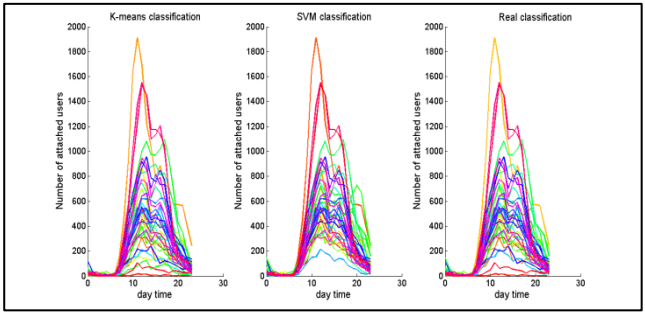

Fig 5. b. Classification result of the morning peak load class, by k-means (left), by SVM (middle) and real classification (right)

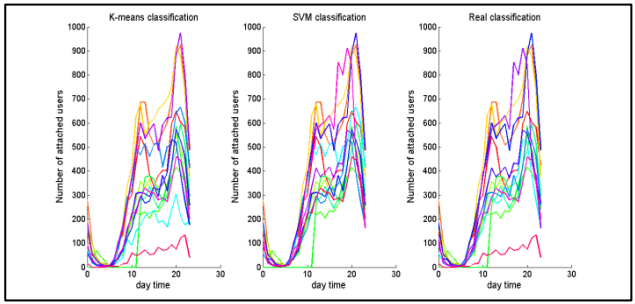

Fig 5. c. Classification result of the evening peak load class, by k-means (left), by SVM (middle) and real classification (right).

TABLE I. COMPARISON OF THE RESULTS OBTAINED FROM SVM AND K-MEANS AGAINST REAL CLASSIFICATION.

| correctly classified | Class 1 | Class 2 | Class 3 | Total |
|---|---|---|---|---|
| SVM | 82.35% | 87.17% | 92.85% | 87.14% |
| Kmeans | 76.47% | 89% | 85.7% | 85% |

Note that we used for k-means an hourly load in the day instead of a granularity of ten minutes as for SVM. Indeed, with k-means, when calculating the distance between two base station vectors, the smaller the time granularity, the higher the noise and the calculation of the distance between both vectors becomes very sensitive to this noise, leading to false classifications. Whereas for SVM, calculations are insensitive to the level of granularity and we can almost obtain the same result in the two cases.

We notice that k-means results are close to SVM classification results and real ones. So, we can as well use k-means tool to form the training set for SVM instead of preparing it by ourselves. In other hand, SVM results can be enhanced by increasing the number of training set, in fact, the bigger the data set size is, the more accurate the result is. However, we cannot use K-means to classify a big set of traces because it uses all the set to perform classification task and it will consume more time processing and more computational requirements, unlike SVM which is more suitable for the task and deals better with big data classification and this is the strong point of SVM.

## V. PREDICTION OF BASE STATION LOAD

### A. Motivations:

We claim that current operators' policies for radio resource allocation and management are no more efficient. The classical technique for allocation is based on a static sharing of the licensed spectrum. Therefore and with the help of our field classifications, we can say that this allocation seems to be a waste for the radio resource as some base stations may not consume all the allocated resource because the low number of attached users. In the same time, there can be some base stations where the number of users is too high and then it demands extra resources to satisfy all the users without affecting their QoS. From our analysis, we notice that the current network planning seems to be not optimal also for energy saving, i.e when there is no user attached to the BS, it still consuming energy uselessly.

Dynamic Techniques for radio resource allocation must substitute the classic ones, so that we can exploit dynamically the unused resources on a given period of time on another places and allocate them to base station when there can be a bottleneck at that moment.

Many techniques can help operators and OEMs (original equipment manufacturers) to change their policy of resource allocation as well as an on-line load prediction or to dynamically add and allocate extra spectrum resources like Tv-whitespace [34]. By predicting the future capacity we can guess the class for which the base station will stand. By this way, we dynamically decide how to allocate resources to this BS and when. In our study, we propose an algorithm based on SVR to predict the load of BS according its history.

### B. Support Vector regression:

SVM algorithm can be applied to classification and prediction issues. Support vector regression (SVR) is modified algorithm of SVM that performs the prediction. SVR is different from conventional regression techniques because it uses Structural Risk Minimization (SRM) but not Empirical Risk Minimization (ERM) induction principle which is equivalent to minimizing an upper bound on the generalization error and not the training error. This feature is expected to perform better than conventional techniques which may suffer from possible over fitting.

## C. Prediction model:

Similarly to the classification model, there are two steps for the prediction model: the training and the testing steps and the principles of those steps are very similar to the previous ones. The difference here concerns the features of the training vectors. For classification, we used the daily BS's load variation which is extracted from the D4D traces dataset.

We propose to define a set of four pertinent features for each base station to predict its load. Each training vector $X_i \in \mathbb{R}^4$ is described as follows:

$$X_i = \{x_i^1, x_i^2, x_i^3, x_i^4\}$$

- Vector $X_i$ corresponds to the BS which we want to predict its capacity at the ith time interval.
- $x_i^1 \in \{1, 2 \ldots 144\}$ stands for the chronological number of the time interval (the duration of each interval is equal to ten minutes) within a day. For example the interval [10:00-10:10] should has the number 61.
- $x_i^2 \in \{1, 2 \ldots 7\}$ represents the number of the corresponding training weekday. We choose this feature to be like that because we noticed that the load can be cyclic.
- $x_i^3 \in \{1, 2 \ldots 52\}$ represents the number of the week.
- $x_i^4 \in \{1, 2 \ldots N\}$ corresponds to the year of the training day. N can vary according to the number of year with which to train the prediction model. In our case we have just a one year of traces.

Like the classification model, each vector must be trained with the corresponding label. For the prediction model, the label $y_i$ of the trained vector $X_i$ represents the load of the BS at a determined interval $x_i^1$.

To find the optimal value for parameters C, $\gamma$ (for RBF) and $\varepsilon$ we used the cross validation technique. Then we adjust the SVR machine with these parameters and test the efficiency of the prediction output.

## D. Results:

To evaluate our contribution, we tested with the D4D traces dataset and considered the mean squared error (MSE) as a criterion for the evaluation. The MSE is measured between the predicted BS's load and the real one extracted from traces.

At first, we trained the SVR machine with 3 weeks of data and used the built model to predict the capacity of the next day. Then, we measured at each cross validation iteration the MSE. Fig 6 depicts the variation of the MSE in function of $\gamma$ parameter. The figure shows that the optimal value is $1.4*10^{-3}$. We fix then this parameter and apply cross validation for C and $\varepsilon$ parameters similarly.

After that, we trained the machine with eleven weeks, with the optimal SVR parameters and we tried to predict the next week. Fig 7 shows the comparison between the real data (in red) and the predicted one (in blue). We notice clearly from the figure that the prediction conserves almost the profile of the BS and predict correctly its class. For example for the week-days, the profile of BS stands for the second class. Despites, on weekend, and especially on Sunday, the profile changes and stands for the first class.

## VI. NETWORK PLANNING TOOL:

Our planning tool is designed to exploit the output of network classification and load prediction algorithms in order to find out an optimal access network topology composed of femtocells which can be turned on/off accordingly to the QoS required and load level. This tool can be integrated to a network simulator to help planner to make planning schemes decision guaranteeing a good QoS for all users. Results proved that our algorithms fit greatly with this need and can be exploited to deliver to the planner a clear vision over his network. Figure 8 presents our tool architecture scheme. The planner must provide as input, the topology parameters required to his simulation scenario like the total number of femtocells, femtocell location, information about the geographic area like its shape files, Qos requirement etc. He must also provide the history of its network traces or CDRs which will be used by SVR as an input in order to forecast the daily load of each deployed femtocell. The output of the SVR machine is used by the SVM to infer the class of each one. This results is also exploited by the planning algorithm box, which communicates also with a femtocell database (contains all information about the installed femtocells), to make the planning decision. This output indicates whether we turn on/off femtocells (turn off if the user demand decrease and reduce energy consumption, or turn on more femtocells where the demand increases so that we add more resources and guarantee a better QoS); the output in this case is used as input for another framework which its signaling flow is described in [35]; or to add extra tv-whitespace radio resources [34] if needed in some location.

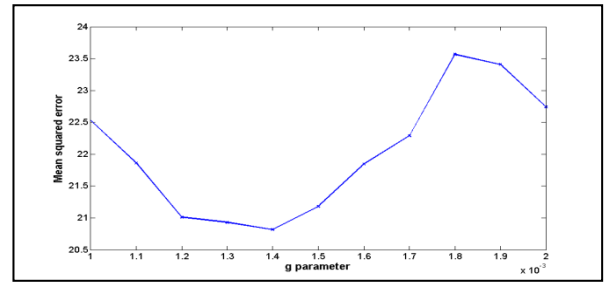

Fig. 6. Variation of MSE in function of γ parameter

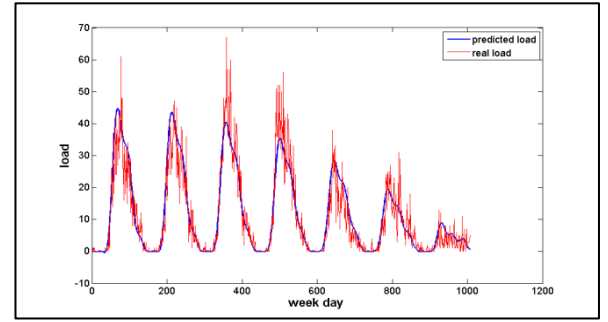

Fig. 7. Prediction results

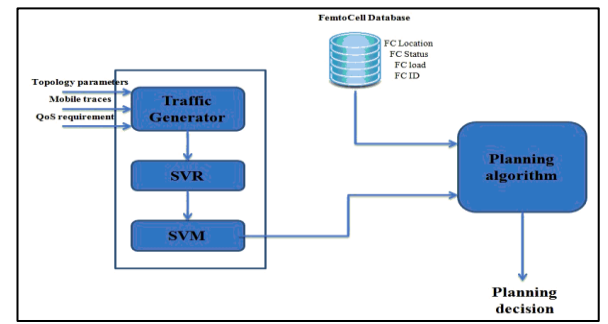

Fig. 8. Network planning tool scheme

## VII. CONCLUSION AND FUTURE WORK

This paper proposes to exploit a real dataset of CDRs and to implement two complementary approaches that aim to help operators to analyze their network, classify traffic and then to predict future behavior. We propose a classification model based on SVM which is capable to group BS into three principal classes according to their daily load profile and we compared it to a classification algorithm based on k-means. Results show that the two classifiers are close with an advantage for SVM which is more adequate for big data classification instead of K-means. We proposed also another algorithm based on SVR to perform an online load prediction for each BS and we obtained excellent results even when the profile changes. All these algorithms are used in a planning tool to help operators to make optimal planning decisions.

Ongoing simulation and modeling work are carried out in order to prove that it is possible to enhance the resource allocation and the quality of service when we exploit these two algorithms. Due to lack of space, more simulations and evaluation results of our planning tool will be presented also on a forthcoming work. We also work on integrating this tool on an open source network simulator to help them to generate realistic optimal topologies taking benefits from our two algorithms. Finally, in the dataset analysis we will study correlation between different base station CDR sets.